\documentclass[runningheads]{llncs}

\usepackage[T1]{fontenc}
\usepackage{amsmath}
\usepackage[table,xcdraw]{xcolor}
\usepackage{multirow}
\usepackage[squaren,Gray]{SIunits}
\def\doi#1{\href{https://doi.org/\detokenize{#1}}{\url{https://doi.org/\detokenize{#1}}}}

\usepackage{graphicx}
\usepackage{listings}
\begin{document}
\title{Beyond Voxel Prediction Uncertainty: Identifying brain lesions you can trust}
\titlerunning{Learning Lesion Uncertainty for 3D MRI Segmentation}

\author{Anonymous}
\authorrunning{Anonymous}
\author{Benjamin Lambert \inst{1, 2} \and
Florence Forbes \inst{3} \and Senan Doyle \inst{2} \and Alan Tucholka \inst{2} \and Michel Dojat \inst{1}}

\authorrunning{B. Lambert et al.}
%
\institute{Univ. Grenoble Alpes, Inserm, U1216, Grenoble Institut Neurosciences, 38000, FR \and
Pixyl, Research and Development Laboratory, 38000 Grenoble, FR \and
Univ. Grenoble Alpes, Inria, CNRS, Grenoble INP, LJK, 38000 Grenoble, FR}
\maketitle              
\begin{abstract}
Deep neural networks have become the gold-standard approach for the automated segmentation of 3D medical images. Their full acceptance by clinicians remains however hampered by the lack of intelligible uncertainty assessment of the provided results. Most approaches to quantify their uncertainty, such as the popular Monte Carlo dropout, restrict to some measure of uncertainty in prediction at the voxel level. In addition not to be clearly related to genuine medical uncertainty, this is not clinically satisfying as most objects of interest (e.g. brain lesions) are made of groups of voxels whose overall relevance may not simply reduce to the sum or mean of their individual uncertainties. In this work, we propose to go beyond voxel-wise assessment using an innovative Graph Neural Network approach, trained from the outputs of a Monte Carlo dropout model. This network allows the fusion of three estimators of voxel uncertainty: entropy, variance, and model's confidence; and can be applied to any lesion, regardless of its shape or size. We demonstrate the superiority of our approach for uncertainty estimate on a task of Multiple Sclerosis lesions segmentation.

\end{abstract}

\section{Introduction}
Magnetic Resonance Imaging (MRI) is the standard imaging modality for the diagnosis and follow-up of Multiple Sclerosis (MS). It allows a direct observation of brain lesions produced by the disease and provides information about the pathology stage or treatment efficiency. Deep Learning (DL) approaches, based on a trained U-Net-like neural network, are invaluable tools to automatically delineate MS lesions \cite{shoeibi2021applications}. Although powerful and versatile, these models provide segmentation maps that are typically opaque, with no indication regarding their certainty. This hinders full acceptance of DL models in clinical routine, for which uncertainty attached to the computerized results is essential for their interpretation and to avoid misleading predictions. 

A variety of methods have been proposed to quantify the uncertainty attached to deep neural networks \cite{abdar2021review}. Among them, the Monte Carlo (MC) dropout stands out as one of the simplest approach, as it can be applied to any model trained with the dropout technique \cite{srivastava2014dropout}. Such a model can be interpreted as a Bayesian neural network, giving access to the interesting properties of these probabilistic models regarding quantification of their uncertainty \cite{gal2016dropout}. More particularly at inference, for a given input, multiple stochastic forward passes are computed by keeping dropout activated, corresponding to empirical samples from the approximated predictive distribution. This produces a set of softmax probabilities that can further be used to compute uncertainty estimates. Applied to MRI segmentation, the MC dropout method produces uncertainty metrics for each voxel in the volume, resulting in so-called voxel-wise uncertainty maps \cite{sander2019towards,jungo2017towards,nair2020exploring}. 
The clinically-relevant information, however, is at a higher level, typically at the instance (lesion, tissue) level.

Natural ways to obtain such instance-wise uncertainties, meaning the uncertainties attached to each connected component within the output segmentation, are through a \textit{post hoc} aggregation  of voxel-wise uncertainty estimations.
Existing approaches include computing the mean uncertainty of voxels belonging to the same class in the segmentation \cite{roy2019bayesian} (thus producing one uncertainty estimate per class, rather than per connected component). In the context of MS, lesion-wise uncertainty was also estimated using the logsum of the connected voxels uncertainties \cite{nair2020exploring}. Using the mean implies that each component uncertainty contributes equally to the overall instance score, while the use of the logsum assumes that connected voxels are conditionally independent, given that they belong to the same instance. These highly simplified assumptions may degrade the quality of instance uncertainty computation. To go further, a side-learner called MetaSeg has been proposed to predict the Intersection Over Unions (IoU) of each individual segmented instance with the ground truth \cite{rottmann2020prediction}. For this task, a Linear Regression Model is trained based on a series of features derived from a standard segmentation model’s output probabilities. The predicted score is then used as a marker of instance uncertainty. Yet, the input features of MetaSeg consist in averaged voxel-wise metrics, leading to the same restrictions than the previously-described \textit{post hoc} aggregation methods. 
Additionally, it has been proposed to train an auxiliary Graph (Convolutional) Neural Network (GCNN) using the outputs of a MC dropout U-Net (i.e. voxel-wise segmentation and uncertainty maps) to refine the predicted masks \cite{soberanis2020uncertainty}. This approach, however, remains at the voxel level and focuses on 2D segmentation tasks. 

In this work, we propose to build from the two last methods to overcome their respective limitations. Indeed, we implement a GCNN at the output of a trained MC dropout U-Net model. Using the predicted 3D segmentation outputs, each individual segmented lesion is modeled by a graph whose voxels are the interconnected nodes. Node features are determined by the input and output of the U-Net, comprising the voxel image intensities, the voxel predicted label, and voxel-wise uncertainty maps. We implement two alternative variants of the proposed GCNN, either classification or regression, to quantify lesions uncertainty. We test our framework on a task of 3D binary segmentation segmentation on MS data. Results demonstrate the superiority of our approach compared to known methods. 

\section{Our Framework: Graph modelization for lesion uncertainty quantification}
\begin{figure}[!htb]
\centering
\includegraphics[width=\textwidth]{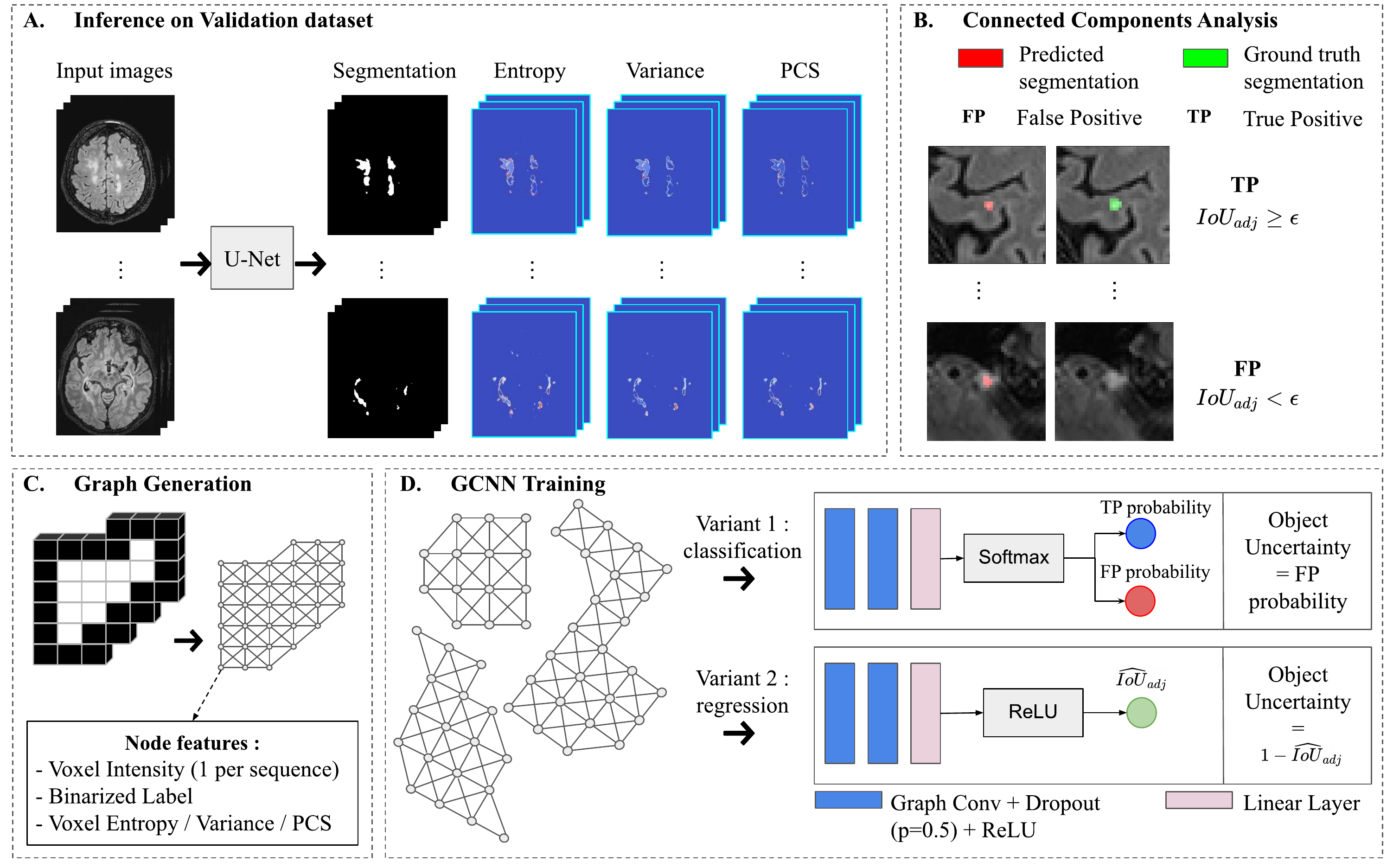}
\caption{Illustration of the proposed framework for learning lesion uncertainty from the outputs of a Monte Carlo dropout model. See the text for details of each block.} \label{framework}
\end{figure}

\textit{Overview:} Consider an input image X and a trained MC dropout segmentation model $\mathcal{N}$ with parameters $W$ that produces a segmentation $Y=\mathcal{N}(X, W)$ and a set of voxel-wise uncertainty maps $U_i$ (e.g. entropy, variance, PCS, etc.). Our objective is to quantify the uncertainty of each instance (i.e.\ lesion) in Y. To do so, we propose to train an auxiliary GCNN to predict this uncertainty directly from X, Y, and $U_i$ (see Figure \ref{framework}). 

\subsection{Monte Carlo dropout model and voxel-wise uncertainty}

We use a generic 3D U-Net \cite{cciccek20163d} for its simplicity and popularity within the field, although our method can be employed with any segmentation model trained with dropout. We add 3D dropout \cite{tompson2015efficient} with a rate of $p=0.2$ at the end of each encoding and decoding block. The model is trained on annotated datasets composed of pairs of images: (i) input T2-weighted FLAIR MRI sequences $X$ and (ii) associated ground truth MS lesions segmentation $Y$. At inference, dropout is kept activated and $T$ forward passes are made for a new input volume $x^*$. We chose $T=20$, as it allows an optimal counterpart between inference time and quality of uncertainty estimates \cite{orlando2019u2}. From this set of predictions, several well-known voxel-related uncertainty metrics are extracted: (see Figure \ref{framework}, part A): the entropy \cite{gal2017deep}, the variance \cite{kendall2015bayesian} and the Predicted Confidence Score (PCS) \cite{zhang2020towards}. 

\subsection{Graph dataset generation}
\subsubsection{Inference on Validation Dataset and Connected Component Analysis.}
After training, the MC dropout U-Net is subsequently used to generate segmentation and uncertainty maps on the set-aside validation set of images. These predictions are used to generate training data for the auxiliary GCNN. We use a Connected Component Analysis (CCA) to identify each lesion in the segmentation masks using 26-connectivity --- meaning that a lesion is defined by voxels that are interconnected by their faces, edges, or corner. For each lesion identified by CCA, we compute the Adjusted Intersection Over Union ($IoU_{adj}$) \cite{rottmann2020prediction} with the ground truth lesions (see Figure \ref{framework}, part B). This variant of the IoU is suited for brain-abnormalities segmentation, where a connected component in the ground truth can be divided into several pieces in the predicted segmentation. 

Identified lesions can exhibit a wide range of shape and size. To learn from these data, we must thus design a neural network that can be employed regardless of the shape and size of the input structure. GCNNs, which can be interpreted as a generalization of the classic convolutional networks to non-Euclidean and irregular data, are thus particularly suitable for this task.

\subsubsection{From voxels to graphs}
We first slightly dilate each lesion mask to include surrounding voxels at the border between classes, which typically convey useful information about uncertainty. We then convert the dilated mask into a graph by representing its voxels by nodes and neighborhood relationships by edges. Each node is further defined by a set of $n+4$ features: (i) the intensity of its corresponding voxel in each of the $n$ input MRI sequences, (ii) its binarized label (1 for the observed lesion class and 0 for all other classes), and its 3 voxel-wise uncertainty estimates: (iii) entropy, (iv) variance and (v) PCS (see Figure \ref{framework}, part C). In agreement with the aforementioned  26-connectivity CCA,  each node (i.e. voxel) is connected in the graph to its 26 nearest neighbors. 
 
\subsection{GCNN architecture and training}
Here, we use a lightweight GCNN architecture composed of 2 consecutive Graph Convolutional layers with a hidden dimension of $h=64$, followed by a Linear layer (see Figure \ref{framework}, part D).
The model is trained using the graph dataset generated from the validation images, composed of graphs (transformed connected components obtained from the segmentation model) along with their associated ground truth ($IoU_{adj}$). As in \cite{rottmann2020prediction}, we propose two versions of our model: 
\begin{itemize}
    \item In the classification approach ($\text{GCNN}_{\text{Classif}}$), the $IoU_{adj}$ labels are first binarized as follows: FP if $IoU_{adj}(graph) < \epsilon$, and TP if $IoU_{adj}(graph) \geq \epsilon$. $\epsilon$ is a hyperparameter that we set to $0.1$ in our experiments, so that lesions with an $IoU_{adj}$ very close to 0 are not wrongly considered as TP. The network is then trained using the Cross-Entropy Loss. At inference, structural uncertainty is quantified by the graph FP probability.
    \item In the regression approach ($\text{GCNN}_{\text{Reg}}$), the model is directly trained to predict the graph $\widehat{IoU}_{adj}$, using the MSE loss. At inference, we use $1-\widehat{IoU}_{adj}$ as the structural uncertainty score. 
    \end{itemize}

\section{Material and Method}
\subsection{Data}
We combine two open-source MS datasets: from the University Hospital of Ljubljana (MSLUB) \cite{lesjak2018novel} and from the MICCAI 2016 MS segmentation challenge (MSSEG 2016) \cite{commowick2021multiple}. We thus use 83 manually-annotated 3D T2-FLAIR sequences. Images are resampled to a \unit{1}{\milli\meter} isotropic resolution of $160\times 192 \times 160$ to focus on brain tissues, and intensities are normalized to zero mean and unit variance. We opt for a 4-fold cross-validation scheme due to the limited number of images. In each fold, we put aside $25\%$ of the images for testing. From the remaining images, we use $20\%$ for validation and $80\%$ to train the model. During evaluation, results are averaged over the 4 folds. Due to the limited number of images, we extensively use Data Augmentation to train our models, comprising flipping, rotation, contrast alteration, gaussian noise and blurring.

\subsection{Comparison with known approaches}
To evaluate the relevance of our proposed $\text{GCNN}_{\text{Classif}}$ and $\text{GCNN}_{\text{Reg}}$ approaches, we implement in parallel known approaches to obtain instance uncertainty from the U-Net. We use the mean and logsum of the voxel-wise uncertainty of each lesions, with the 3 different types of uncertainty. We name these methods $\text{Entropy}_{\text{mean}}$, $\text{Variance}_{\text{mean}}$, $\text{PCS}_{\text{mean}}$, $\text{Entropy}_{\text{logsum}}$, $\text{Variance}_{\text{logsum}}$, and $\text{PCS}_{\text{logsum}}$.

As pointed out in \cite{nair2020exploring}, using the logsum assigns a higher uncertainty to small-size lesions. This appears sub-optimal as small lesions could be segmented with high confidence, especially in the case of MS lesions. To verify this point, we implement a naive approach, named Size, which attributes a lesion uncertainty inversely proportional to its size. The lesion size (number of voxels composing it) being $S$, its uncertainty is computed as $1/S$.

Lastly, we implement an approach inspired from the MetaSeg framework \cite{rottmann2020prediction}. We extract a series of features from each connected component in the validation dataset, consisting in the mean entropy, variance and PCS, as well as the size of the lesion. We then train a Logistic Regression classifier from these 4 features to distinguish between True Positive (TP) and FP lesions ($\text{MetaSeg}_{\text{Classif}}$). Alternatively, we train a Linear Regression model to directly predict $\widehat{IoU}_{adj}$ ($\text{MetaSeg}_{\text{Reg}}$). We use the outputs of these models to obtain lesion uncertainty as described for the GCNN approach.

\subsection{Evaluation Setting}
For medical applications, the ideal uncertainty quantification should attribute a higher uncertainty to FP lesions than TP, to allow for proper interpretation and evaluation of the results. To evaluate this properly, we use Accuracy-Confidence curves \cite{lakshminarayanan2017simple}. Briefly, the principle is to set aside the $\tau \%$ of the most uncertain predicted lesions among the test dataset, and measure the performance of the model on the remaining lesions by counting the number of FP and TP lesions. The threshold $\tau$ fluctuates between 0 (all lesions are kept) and 100 (all lesions are removed). By plotting the couples (FP, TP) at different thresholds, we obtain an Accuracy-Confidence curve and compute the AUC (Area Under the Curve) score reflecting the quality of the estimated lesion uncertainty. FP and TP counts are normalized in the range $[0, 1]$ by dividing by the counts obtained without filtering (at $\tau = 0)$. This metric only depends on the ranking of uncertainties, thus is independent of the uncertainty ranges of each method ensuring a fair comparison. We additionally evaluate the segmentation performance of the U-Net on the test datasets using the Dice coefficient, as well as the total number of TP and FP lesions. Finally, for each method, we control the correlation between the estimated uncertainty and the lesion size using the Spearman's rank correlation coefficient ($\rho$).

\begin{table}[!htb]
\centering
\caption{U-Net segmentation performance on the MS dataset and number of TP and FP lesions for each fold.}\label{tab_seg}
\begin{tabular}{|c|c|c|c|c|}
\hline
 Fold & 0 & 1 & 2 & 3  \\ \hline
 Dice & 0.672 & 0.645 & 0.705 & 0.693 \\ \hline
\# TP lesions & 829 & 597 & 715 & 871 \\ \hline
\# FP lesions & 525 & 294 & 353 & 454 \\ \hline
\end{tabular}
\end{table}

\subsection{Implementation Details}
\subsubsection{3D Segmentation U-Net}
Our segmentation framework was implemented using PyTorch \cite{NEURIPS2019_9015}. We opt for a patch approach to train the segmentation U-Net, meaning that the $160\times192\times160$ MRI volumes are split into 3d patches of $160\times 192 \times 32$,  decreasing the memory cost of training. We use a batch size of 5. The U-Net is trained with a combination of the Dice \cite{milletari2016v} and Cross-Entropy loss, using the ADAM optimizer \cite{ADAM} with a learning rate of $1e^{-4}$ until convergence. For the training of the segmentation models, a single NVIDIA T4 GPU was used. 

\subsubsection{Graph Neural Networks}
We use the Deep Graph Library \cite{wang2019deep} to implement and train the GCNN models. The training procedure of our GCNN is standard: we use the ADAM optimizer with a learning rate of $1e^{-2}$ at the start of training, and progressively decreasing to $1e^{-5}$. Graphs are presented to the network in the form of batches, composed of 10 graphs. Due to the small size of the GCNN models, they were trained on CPU, which took a couple of minutes in our experiments. 

\begin{figure}[!htb]
\centering
\includegraphics[width=0.75\textwidth]{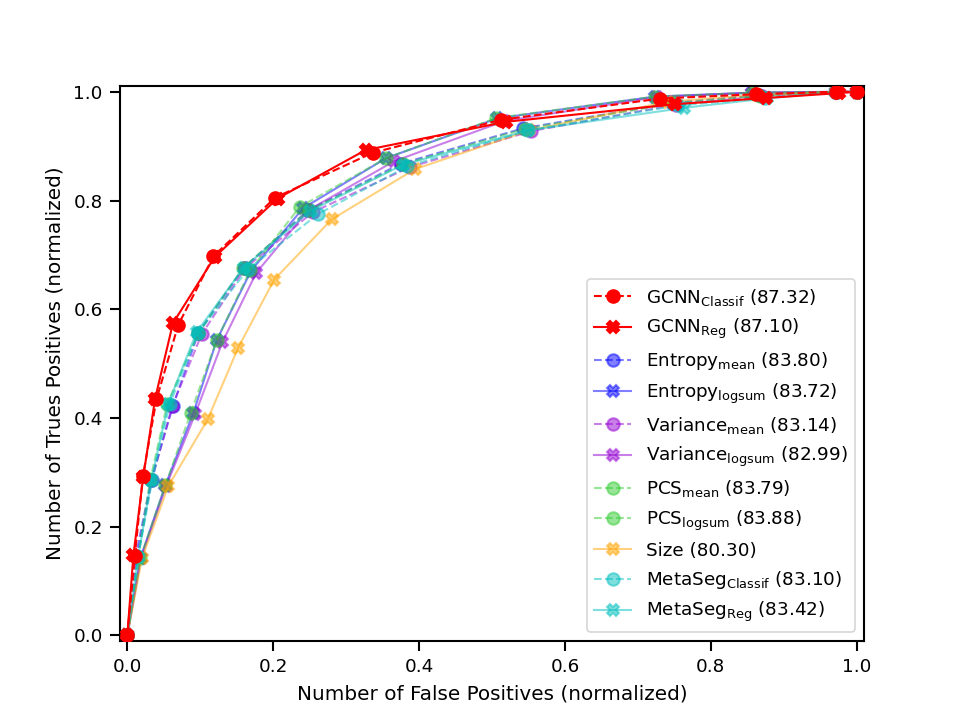}
\caption{Accuracy-Confidence curves for the different methods. The associated AUC scores are indicated in brackets in the graphs legends.} \label{res_auc}
\end{figure}

\section{Results and Discussion}
Accuracy-Confidence curves are presented in Figure \ref{res_auc} along with the corresponding AUC values. Segmentation performance and correlation coefficients are presented in Tables \ref{tab_seg} and \ref{tab_auc}. Experimental results show that both models of the proposed framework outperform the classical methods by a significant margin, and that their performances are similar with a very small advantage for the classification version. The naive Size approach achieves the lowest performance. Similarly, the $logsum$ approaches, also strongly correlated with the lesion size, have poorer performance than the $mean$ counterparts.  Not surprisingly, in the context of MS lesions, the lesion size is not a satisfying proxy for uncertainty as small lesions can be segmented with high confidence. In our experimental setting, $MetaSeg$ models do not outperform simpler methods. This is probably due to the overall simplicity of these models, failing to fully learn the relationships between the different input features.

Results show that our graph-based framework can be efficiently used to flag uncertain lesions, that are more likely to result in False Positives. The classification variant slightly outperforms regression. We hypothesize that this is due to the increased difficulty of predicting the exact $IoU_{adj}$, compared to the binary classification setting. One drawback of our approach is that it requires an additional validation set containing enough lesions (typically a few hundreds) to allow GCNN training. However, as most DL pipelines rely on a set-aside validation set to control overfitting during training, these data can then be used for this purpose (as it was the case in this work). The requirement is thus not prohibitive and only necessitates a sufficiently large validation set. 

Overall, our framework is computationally light as CCA is computed only once per MRI, followed by the graph generation step that can be parallelized among the lesions. Additionally, in the context of MS, most brain lesions are relatively small (less than 1000 voxels), which results in small graphs that are fast to generate. Finally we use 26-connectivity, meaning that a voxel is only connected to its closest neighbors, which reduces the computational burden.  

Our approach enhances the binary voxel-wise predictions of the segmentation model with reliable and readable lesion-wise uncertainty estimates. In the classification setting, uncertainty is cast as the probability of a lesion being a false positive, which is a straightforward and intelligible definition. In a real world clinical application, this may help the clinician examine the automated segmentation in the light of the model's confidence, hence allowing a better interpretability of the provided results and a more trustable usage of the algorithm. 

Future work will study the extension to multi-class segmentation, and inclusion of additional features such as the global location of the lesion within the MRI volume. Indeed, for brain disorders such as MS, the location of the lesion within the brain conveys information concerning uncertainty, as false positives tend to be concentrated in specific brain regions. 

\begin{table}[h!]
\centering
\caption{Evaluation of uncertainty estimates (AUC values). $\rho$ represents Spearman's rank correlation coefficient $\rho$. }\label{tab_auc}
\begin{tabular}{l|cc|}
\cline{2-3}
 & \multicolumn{1}{c|}{AUC (\%)} & Spearman's $\rho$ \\ \hline
 \multicolumn{1}{|l|}{$\text{GCNN}_{\text{Classif}}$} & \multicolumn{1}{c|}{\textbf{87.32}} & -0.78 \\ \hline
\multicolumn{1}{|l|}{$\text{GCNN}_{\text{Reg}}$} & \multicolumn{1}{c|}{87.10} & -0.77 \\ \hline
\multicolumn{1}{|l|}{$\text{Entropy}_{\text{mean}}$} & \multicolumn{1}{c|}{83.80} & -0.42 \\ \hline
\multicolumn{1}{|l|}{$\text{Entropy}_{\text{logsum}}$} & \multicolumn{1}{c|}{83.72} & -0.97 \\ \hline
\multicolumn{1}{|l|}{$\text{Variance}_{\text{mean}}$} & \multicolumn{1}{c|}{83.14} & -0.44 \\ \hline
\multicolumn{1}{|l|}{$\text{Variance}_{\text{logsum}}$} & \multicolumn{1}{c|}{82.99} & -0.99\\ \hline
\multicolumn{1}{|l|}{$\text{PCS}_{\text{mean}}$} & \multicolumn{1}{c|}{83.79} & -0.44 \\ \hline
\multicolumn{1}{|l|}{$\text{PCS}_{\text{logsum}}$} & \multicolumn{1}{c|}{83.88} & -0.98 \\ \hline
\multicolumn{1}{|l|}{Size} & \multicolumn{1}{c|}{80.30} & -1.00  \\ \hline
\multicolumn{1}{|l|}{$\text{MetaSeg}_{\text{Classif}}$} & \multicolumn{1}{c|}{83.10} & -0.76 \\ \hline
\multicolumn{1}{|l|}{$\text{MetaSeg}_{\text{Reg}}$} & \multicolumn{1}{c|}{83.42} & -0.77 \\ \hline
\end{tabular}
\end{table}

\section{Conclusion}
This paper presents an innovative graph-based framework to quantify lesion-wise uncertainty. We demonstrate, with our approach, improvement of the predicted uncertainty, when compared to various known methods. The strength of our solution is its generic nature, making it compatible with any segmentation model trained with dropout.

%
%

\bibliographystyle{splncs04}
\bibliography{biblio}

\begin{thebibliography}{10}
\providecommand{\url}[1]{\texttt{#1}}
\providecommand{\urlprefix}{URL }
\providecommand{\doi}[1]{https://doi.org/#1}

\bibitem{abdar2021review}
Abdar, M., Pourpanah, F., Hussain, S., Rezazadegan, D., Liu, L., Ghavamzadeh,
  M., Fieguth, P., Cao, X., Khosravi, A., Acharya, U.R., et~al.: A review of
  uncertainty quantification in deep learning: Techniques, applications and
  challenges. Information Fusion  \textbf{76},  243--97 (2021)

\bibitem{cciccek20163d}
{\c{C}}i{\c{c}}ek, {\"O}., Abdulkadir, A., Lienkamp, S.S., Brox, T.,
  Ronneberger, O.: {3D U-Net: learning dense volumetric segmentation from
  sparse annotation}. In: International conference on medical image computing
  and computer-assisted intervention. pp. 424--432. Springer (2016)

\bibitem{commowick2021multiple}
Commowick, O., Kain, M., Casey, R., Ameli, R., Ferr{\'e}, J.C., Kerbrat, A.,
  Tourdias, T., Cervenansky, F., Camarasu-Pop, S., Glatard, T., et~al.:
  {Multiple sclerosis lesions segmentation from multiple experts: The MICCAI
  2016 challenge dataset}. NeuroImage  \textbf{244},  118589 (2021)

\bibitem{gal2016dropout}
Gal, Y., Ghahramani, Z.: {Dropout as a Bayesian approximation: Representing
  model uncertainty in deep learning}. In: international conference on machine
  learning. pp. 1050--1059. PMLR (2016)

\bibitem{gal2017deep}
Gal, Y., Islam, R., Ghahramani, Z.: {Deep Bayesian active learning with image
  data}. In: International Conference on Machine Learning. pp. 1183--1192. PMLR
  (2017)

\bibitem{jungo2017towards}
Jungo, A., McKinley, R., Meier, R., Knecht, U., Vera, L., P{\'e}rez-Beteta, J.,
  Molina-Garc{\'\i}a, D., P{\'e}rez-Garc{\'\i}a, V.M., Wiest, R., Reyes, M.:
  Towards uncertainty-assisted brain tumor segmentation and survival
  prediction. In: International MICCAI Brainlesion Workshop. pp. 474--485.
  Springer (2017)

\bibitem{kendall2015bayesian}
Kendall, A., Badrinarayanan, V., Cipolla, R.: {Bayesian SegNet: Model
  Uncertainty in Deep Convolutional Encoder-Decoder Architectures for Scene
  Understanding}. In: British Machine Vision Conference 2017, {BMVC} 2017,
  London, UK, September 4-7, 2017. {BMVA} Press (2017)

\bibitem{ADAM}
Kingma, D.P., Ba, J.: Adam: {A} method for stochastic optimization. In: Bengio,
  Y., LeCun, Y. (eds.) 3rd International Conference on Learning
  Representations, {ICLR} 2015, San Diego, CA, USA, May 7-9, 2015, Conference
  Track Proceedings (2015), \url{http://arxiv.org/abs/1412.6980}

\bibitem{lakshminarayanan2017simple}
Lakshminarayanan, B., Pritzel, A., Blundell, C.: Simple and scalable predictive
  uncertainty estimation using deep ensembles. In: Advances in Neural
  Information Processing Systems 30: Annual Conference on Neural Information
  Processing Systems 2017. pp. 6402--6413 (2017)

\bibitem{lesjak2018novel}
Lesjak, {\v{Z}}., Galimzianova, A., Koren, A., Lukin, M., Pernu{\v{s}}, F.,
  Likar, B., {\v{S}}piclin, {\v{Z}}.: {A novel public MR image dataset of
  multiple sclerosis patients with lesion segmentations based on multi-rater
  consensus}. Neuroinformatics  \textbf{16}(1),  51--63 (2018)

\bibitem{milletari2016v}
Milletari, F., Navab, N., Ahmadi, S.A.: V-net: Fully convolutional neural
  networks for volumetric medical image segmentation. In: 2016 fourth
  international conference on 3D vision (3DV). pp. 565--571. IEEE (2016)

\bibitem{nair2020exploring}
Nair, T., Precup, D., Arnold, D.L., Arbel, T.: Exploring uncertainty measures
  in deep networks for multiple sclerosis lesion detection and segmentation.
  In: Medical Image Computing and Computer Assisted Intervention - {MICCAI}
  2018 - 21st International Conference, Proceedings, Part {I}. Lecture Notes in
  Computer Science, vol. 11070, pp. 655--663. Springer (2018)

\bibitem{orlando2019u2}
Orlando, J.I., Seeb{\"o}ck, P., Bogunovi{\'c}, H., Klimscha, S., Grechenig, C.,
  Waldstein, S., Gerendas, B.S., Schmidt-Erfurth, U.: U2-net: A bayesian u-net
  model with epistemic uncertainty feedback for photoreceptor layer
  segmentation in pathological oct scans. In: 2019 IEEE 16th International
  Symposium on Biomedical Imaging (ISBI 2019). pp. 1441--1445. IEEE (2019)

\bibitem{NEURIPS2019_9015}
Paszke, A., Gross, S., Massa, F., Lerer, A., Bradbury, J., Chanan, G., Killeen,
  T., Lin, Z., Gimelshein, N., Antiga, L., Desmaison, A., Kopf, A., Yang, E.,
  DeVito, Z., Raison, M., Tejani, A., Chilamkurthy, S., Steiner, B., Fang, L.,
  Bai, J., Chintala, S.: Pytorch: An imperative style, high-performance deep
  learning library. In: Advances in Neural Information Processing Systems 32,
  pp. 8024--8035. Curran Associates, Inc. (2019)

\bibitem{rottmann2020prediction}
Rottmann, M., Colling, P., Hack, T.P., Chan, R., H{\"u}ger, F., Schlicht, P.,
  Gottschalk, H.: Prediction error meta classification in semantic
  segmentation: Detection via aggregated dispersion measures of softmax
  probabilities. In: 2020 International Joint Conference on Neural Networks
  (IJCNN). pp.~1--9. IEEE (2020)

\bibitem{roy2019bayesian}
Roy, A.G., Conjeti, S., Navab, N., Wachinger, C., Initiative, A.D.N., et~al.:
  {Bayesian QuickNAT: Model uncertainty in deep whole-brain segmentation for
  structure-wise quality control}. NeuroImage  \textbf{195},  11--22 (2019)

\bibitem{sander2019towards}
Sander, J., de~Vos, B.D., Wolterink, J.M., I{\v{s}}gum, I.: {Towards increased
  trustworthiness of deep learning segmentation methods on cardiac MRI}. In:
  Medical Imaging 2019: Image Processing. vol. 10949, p. 1094919. International
  Society for Optics and Photonics (2019)

\bibitem{shoeibi2021applications}
Shoeibi, A., Khodatars, M., Jafari, M., Moridian, P., Rezaei, M., Alizadehsani,
  R., Khozeimeh, F., Gorriz, J.M., Heras, J., Panahiazar, M., et~al.:
  Applications of deep learning techniques for automated multiple sclerosis
  detection using magnetic resonance imaging: A review. Computers in Biology
  and Medicine  \textbf{136},  104697 (2021)

\bibitem{soberanis2020uncertainty}
Soberanis-Mukul, R.D., Navab, N., Albarqouni, S.: Uncertainty-based graph
  convolutional networks for organ segmentation refinement. In: Medical Imaging
  with Deep Learning. pp. 755--769. PMLR (2020)

\bibitem{srivastava2014dropout}
Srivastava, N., Hinton, G., Krizhevsky, A., Sutskever, I., Salakhutdinov, R.:
  Dropout: a simple way to prevent neural networks from overfitting. The
  journal of machine learning research  \textbf{15}(1),  1929--1958 (2014)

\bibitem{tompson2015efficient}
Tompson, J., Goroshin, R., Jain, A., LeCun, Y., Bregler, C.: Efficient object
  localization using convolutional networks. In: Proceedings of the IEEE
  conference on computer vision and pattern recognition. pp. 648--656 (2015)

\bibitem{wang2019deep}
Wang, M., Zheng, D., Ye, Z., Gan, Q., Li, M., Song, X., Zhou, J., Ma, C., Yu,
  L., Gai, Y., et~al.: Deep graph library: A graph-centric, highly-performant
  package for graph neural networks. arXiv preprint arXiv:1909.01315  (2019)

\bibitem{zhang2020towards}
Zhang, X., Xie, X., Ma, L., Du, X., Hu, Q., Liu, Y., Zhao, J., Sun, M.: Towards
  characterizing adversarial defects of deep learning software from the lens of
  uncertainty. In: 2020 IEEE/ACM 42nd International Conference on Software
  Engineering (ICSE). pp. 739--751. IEEE (2020)

\end{thebibliography}

\end{document}